\documentstyle[aps,floats,graphicx,amssymb]{revtex}
\begin{document}
\draft
\wideabs{
\title{EXAFS indication of double-well potential for oxygen vibration in
Ba$_{1-x}$K$_x$BiO$_3$}
\author{A. P. Menushenkov and K. V. Klementev}
\address{Moscow State Engineering Physics Institute, 115409 Moscow, Russia}
\date{\today}
\maketitle
\begin{abstract}
X-ray absorption spectra of oxide systems Ba$_{1-x}$K$_x$BiO$_3$ and
BaPbO$_3$ above Bi- and Pb-$L_3$ absorption edges were investigated. It was
shown that oxygen ions move in double-well potential and their oscillations are
correlated with charge carrier movement. Observed breathing-like oxygen
vibration in double-well potential with large amplitude and low frequency
causes the strong electron-phonon coupling and high $T_c$ values in doped
BaBiO$_3$. Based on the experimental data, the model of relationship of
electronic and local crystal structures is proposed that is in a good agreement
with the results of transport measurements, inelastic neutron and electron
scattering, Raman scattering, and photoemission spectroscopy. In the framework
of the model the possible reasons of superconductivity in perovskite-like
oxides are discussed.
\end{abstract}
}

\section{Introduction}\label{Intr}
Though superconductivity in BaPb$_{1-x}$Bi$_x$O$_3$ (BPBO) was discovered
significantly earlier \cite{Sleight1} than in cuprates, the question of the
nature of superconducting state in this oxide as well as in cognate system
Ba$_{1-x}$K$_x$BiO$_3$ (BKBO) is still unsolved.

The structures of crystal lattice of copper oxide high temperature
superconductors (HTSC's) and bismuth-based oxides have some important common
characteristics. The both oxide classes have perovskite-like lattice
with CuO$_n$ ($n$=4, 5, 6) or Bi(Pb)O$_6$ complexes joined by the common oxygen
ions. In bismuthates, the intersection of octahedral complexes in the three
crystallographic directions determines their three-dimensional cubic structure.
The CuO$_n$ complexes are joined in CuO$_2$ planes, which makes the
two-dimensional structure of copper-oxides.

Because of strong hybridisation of covalent Bi(Pb)$6s$, Cu$3d$ -- O$2p_\sigma$
bonds, the above mentioned complexes are the most tightly bound items of the
perovskite-like structure. Therefore such important peculiarities of
perovskite structure as lattice instability in respect to soft tilting mode of
CuO$_n$ or BiO$_6$ complexes (see for review \cite{Plakida2}) and
highly anisotropic thermal factors of oxygen ions vibration
\cite{Wignacourt_Kwei}, which point out the large amplitude of rotation
oscillations, are inherent to the both classes of superconducting oxides and
cause anharmonic vibrations of oxygen atoms that may be described by movement
in a double-well potential \cite{Plakida1,Hardy1}. These structural
instabilities of perovskite-like lattice can be related with the transition to
superconducting state \cite{Plakida1,Hardy1,Plakida2}.

The layered structure of copper oxide compounds, presence of several
non-equivalent copper positions, and a number of different Cu-O bonds
complicate the local structure analysis. At the same time, the simplicity of
cubic three-dimensional structure of BPBO-BKBO systems makes the interpretation
of experimental data easier to a great extent. Relatively low temperatures of
superconducting transition $T_c\simeq13$\,K in BaPb$_{0.75}$Bi$_{0.25}$O$_3$
\cite{Sleight1} and $T_c\simeq30$\,K in Ba$_{0.6}$K$_{0.4}$BiO$_3$
\cite{Mattheiss3}, the values of superconducting gap
$2\Delta(0)/kT_c=3.6\pm0.1$ for BPBO \cite{Stepankin1} and
$2\Delta(0)/kT_c=3.5\pm0.5$ for BKBO \cite{Schlesinger1}, and a sizeable oxygen
isotope effect \cite{Allen1} allow one to rely on standard BCS-theory of
superconductivity, not excluding a possible realisation of other mechanisms.
The simpler electronic structure of $s-p$ valence band of BPBO-BKBO systems in
comparison with $d-p$ band of cuprates favours the establishment of relationship
of crystal and electronic structures in these compounds.

However, even for these relatively simple systems there is no agreement so
far on a number of crucial aspects: crystal structure symmetry and lattice
dynamics \cite{Marx1,Pei2}, electronic structure \cite{Mattheiss1},
bismuth valence state \cite{Cox2,Wertheim1,Akhtar1,Hashimoto2}. The most part of unusual
properties of BaBiO$_3$-family compounds mentioned in early review
\cite{Uchida1} are still unexplained.

In addition, the average structural data contradicts the local ones. Integral
crystallographic methods point to the simple cubic lattice in BKBO ($x>0.37$)
\cite{Pei2}. In contrast to this, the EXAFS-analysis of four nearest spheres of
bismuth environment \cite{Yacoby1} reveals a local tilting of octahedra by
4--5$^\circ$, and Raman spectra evidence for a local lowering of symmetry from
simple cubic \cite{Anshukova1}.

EXAFS-analysis of the nearest oxygen octahedral environment of Bi in metallic
BKBO compound with $x=0.4$ previously was made in harmonic approach
\cite{Heald1,Salem1} and pointed out at least one unresolved problem.
Temperature dependence of Debye-Waller factor $\sigma^2(T)$ found by Heald {\it
et al.} \cite{Heald1} can be described in Einstein approximation, which is
suitable for harmonic systems, only if one takes into account the temperature
independent factor of 0.0025\,\r{A}$^2$. The attempt to explain this fact due to
some static disorder with doping was denied since $\sigma^2(T)$ of the next
Bi-Ba and Bi-Bi shells show the absence of any disorder. Similar weakly varying
$\sigma^2(T)$ dependence was found in \cite{Salem1}. Thus, it was
observed that amplitude of low temperature Bi-O vibrations in metallic BKBO
compound with $x=0.4$ is too high (more than twice larger than in metallic
BaPbO$_3$ compound) \cite{Boyce1,OurPaper4}.

Besides, our careful measurements \cite{OurPaper4} of temperature dependent
EXAFS-spectra of undoped BaBiO$_3$ which were also treated in harmonic
approach showed absolutely abnormal increasing of $\sigma^2$ values of
both Bi-O bonds with temperature decreasing from 90\,K. We explained these
anomalies by the influence of anharmonic rotational vibrations of oxygen
octahedra which become pronounced at low temperatures \cite{Koyama2}, the model
of connection of superconductivity with rotation mode asymmetry was proposed
\cite{Menushenkov1_4}.

The above contradictions in description for local structure lead us to
necessity of EXAFS analysis of the nearest Bi-O shell in anharmonic approach.

In the present paper we report the results of temperature dependent
EXAFS-investigation of BPBO-BKBO systems firstly treated in anharmonic
approach based on the idea of such an analysis for apical oxygen atoms in
YBa$_2$Cu$_3$O$_{7-\delta}$ proposed by J.~Mustre de~Leon {\it et al.}
\cite{deLeon1} We have analysed EXAFS function $\chi(k)$, using the new program
``VIPER for Windows'' \cite{OurConf12}, by construction of the model potentials
of atomic vibrations, subsequent calculation of the pair radial distribution
function, and calculation of the model $\chi(k)$. This new approach for the
EXAFS-analysis of Bi-based oxides gives us an opportunity to investigate the
character of oxygen atom vibrations and, in combination with our previous local
electronic structure studies by XANES (x-ray absorption near edge structure)
spectroscopy \cite{Ignatov1}, to understand the nature of the structural phase
transitions in these systems and to explain practically all above
contradictions.

In Sec.~\ref{experiment} we describe the experimental details and present the
procedure of data treatment. The general results are given in
Sec.~\ref{results}. In Sec.~\ref{Relationship} we relate the local crystal
structure with local electronic structure, which is the base of our processing
of experimental data. In Sec.~\ref{supercon} we discuss the possible
superconductive mechanism and demonstrate how the oscillations of oxygen atoms
in double-well potential contributes to superconductivity in BKBO.

\section{Experimental and data analysis}\label{experiment}

\begin{figure}[!t]\begin{center}\includegraphics*[width=\hsize]{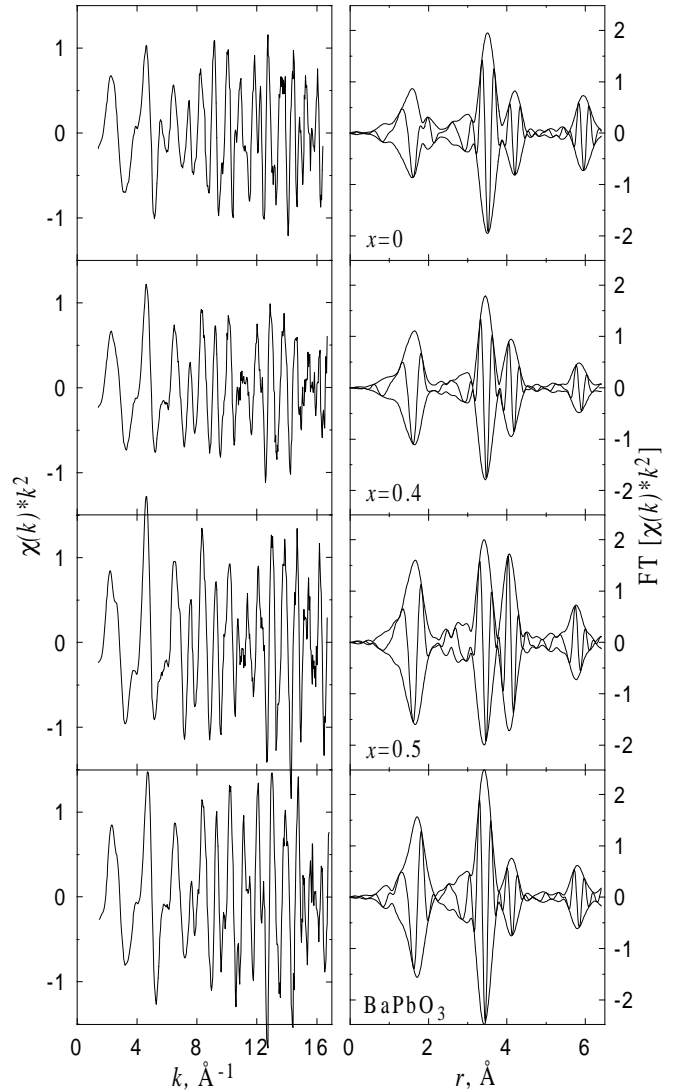}
\end{center}\caption{Experimental EXAFS $\chi(k)k^2$ (left) and its Fourier
transform magnitude and imaginary part (right) for Ba$_{1-x}$K$_x$BiO$_3$ ($x=$
0, 0.4, 0.5) and BaPbO$_3$ at 7\,K.} \label{menush01}\end{figure}

In this work we have investigated ceramic samples of BaPbO$_3$ and BKBO with
$x=$ 0, 0.4, 0.5 synthesised as described in \cite{Anshukova2}. The
materials were examined by x-ray powder diffraction for phase purity. The
samples were controlled by transport and susceptibility measurements. BaBiO$_3$
showed semiconductor-like behaviour, the compositions BKBO with $x=0.4$ and 0.5
showed typical metallic $\rho(T)$ dependence and superconducting properties
with $T_c\simeq30$\,K and $T_c\simeq16$\,K, correspondingly. BaPbO$_3$ samples
were metallic but not superconducting at all temperatures.

For the XAS measurements, a crushed fine powder was precipitated onto a
micropore substrate. The thickness of samples was about two absorption
lengths at the chosen absorption edge.

The x-ray absorption spectra were collected at D-21 line (XAS-13) of DCI
(LURE,Orsay, France) synchrotron operated at energy 1.85 GeV and the average
current $\sim250$\,mA of positron beam at the $L_3$ edges of Bi (13040.6 eV)
and Pb (13426 eV). Energy resolution of the double-crystal Si [311]
monochromator (detuned to reject 50\% of the incident signal in order to
minimise harmonic contamination) with a 0.4\,mm slit at 13 keV was about
2--3\,eV. The low temperature measurements were carried out using a liquid
helium circulation type cryostat with a temperature control of $\pm1$\,K.

The background in the experimental spectra was removed as described in
\cite{Newville1}, taking care to remove the low frequency
oscillations. EXAFS-functions $\chi(k)k^2$ obtained from absorption spectra
were Fourier transformed in the wave number range $k$ from 1.5 to
16.5\,\r{A}$^{-1}$, using Kaiser-Bessel windowing function. Back Fourier
transform was done using a Hanning window from $\sim1$ to $\sim2$\,\r{A}
corresponding to the first Bi-O near-neighbour shell. In such a case, the number
of independent experimental points \cite{Stern1} was
$N_{exp}=2\Delta k\Delta r/\pi+2\approx11$.

The model EXAFS-function for pair atomic absorber-scatterer oscillations is
constructed as follows. Suppose we know the potential of these oscillations as
a parametric function of interatomic distance. Solving the stationary
Sr\"{o}dinger equation for the particle with reduced mass of the atomic pair
\cite{OurConf12}, one obtains a pair radial distribution function (PRDF) of
atoms in $i$-th sphere:
\begin{equation}\label{PRDF}
g_i(r)=N_i\sum_n|\Psi_n(r)|^2e^{-E_n/kT}/\sum_n e^{-E_n/kT},
\end{equation}
where $N_i$ is the coordination number, $E_n$ and $\Psi_n$ are $n$-th energy
level and its corresponding wave function. Given PRDF's, the model EXAFS
function calculated as
\begin{equation}\label{gChi}
\chi(k)=\frac{1}{k}\sum_i
F_i(k)\!\int\limits_{r_{min}}^{r_{max}}\!\!g_i(r)\sin[2kr+\phi_i(k)]/r^2 \, dr,
\end{equation}
where $k=\sqrt{2m_e/\hbar^2(E-E_{th})}$ is the photoelectron wave number
referenced to the ionisation threshold $E_{th}$, $r_{min}$ and $r_{max}$ are
determined by the windowing function of back Fourier transform. The phase shift
$\phi_i(k)$ and scattering amplitude $F_i(k)$ were calculated using FEFF-6 code
\cite{FEFF} for six-shell cluster with crystallographic data from neutron
diffraction study \cite{Pei2} and using default set of the FEFF-6 parameters.
The potential parameters were extracted from the model-to-experimental
EXAFS-function fits.

\section{General results}\label{results}

In Fig.~\ref{menush01} (left) we show the experimental
$\chi(k)k^2$ for BKBO x=0, 0.4, 0.5 and for BaPbO$_3$ measured at $L_3$ Bi(Pb)
absorption edge at 7\,K. Good signal-to-noise ratio seen even for maximal
wave number values $k\gtrsim16$\,\r{A}$^{-1}$ indicates the high spectra
quality. The absence of signal on the Fourier transform in the low-$r$ range
Fig.~\ref{menush01} (right) testifies for correct background removal procedure.

\subsection{BaBiO$_3$}

\begin{figure}[!t]\begin{center}\includegraphics*[width=\hsize]{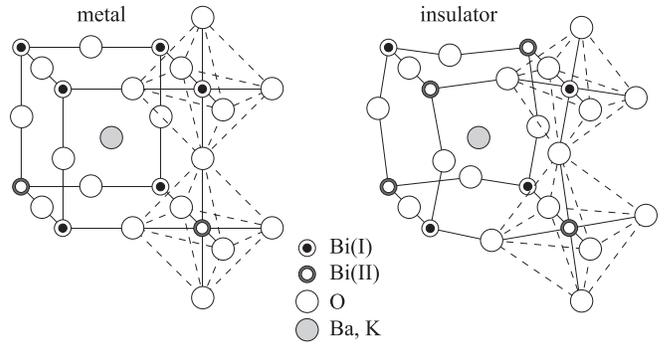}
\end{center}\caption{Sketch of crystal structure of Ba$_{1-x}$K$_x$BiO$_3$ in
metallic ($x>0.37$) and insulating ($x<0.37$) states.}
\label{menush02}\end{figure}

EXAFS-researches \cite{Balzarotti1,Boyce1} of BaBiO$_3$ confirm the results of
crystallographic works \cite{Cox2,Marx1,Pei2}. According to them, there exist
two inequivalent bismuth positions characterised by two Bi-O bond lengths. The
equality of coordination numbers of the two BiO$_6$ spheres points out that the
BaBiO$_3$ structure represents the ordered alternation of small and large
BiO$_6$ octahedra in barium lattice. Such an alternation together with static
rotation distortion around [110] axis produce the monoclinic distortion of
cubic lattice \cite{Cox2,Marx1,Pei2} (Fig.~\ref{menush02}).
As will be shown in Sec.~\ref{Relationship}, to the larger soft octahedra
corresponds the configuration BiO$_6$, and to the smaller rigid octahedra
corresponds Bi\underline{L}$^2$O$_6$. Here, \underline{L}$^2$ denotes the hole
pair in antibonding Bi$6s$O$2p_{\sigma^*}$ orbital of the octahedral complex.

\begin{figure}[!t]\begin{center}\includegraphics*[width=\hsize]{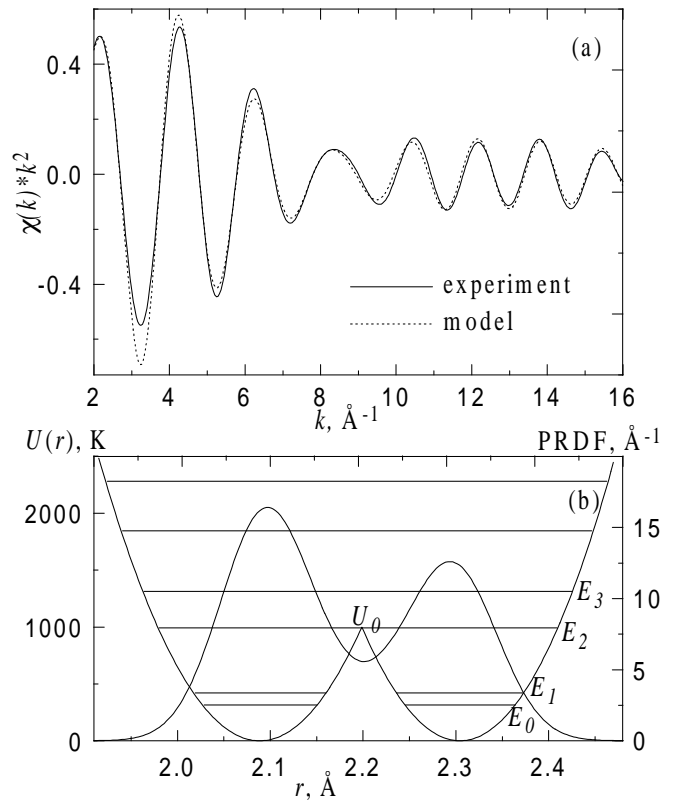}
\end{center}\caption{Experimental and model $\chi(k)k^2$ EXAFS for BaBiO$_3$ at
7\,K for the first Bi-O near-neighbour shell (a) and the model potential with
corresponding PRDF and energy levels (b). $U_0$ is the energy of potential
barrier} \label{menush03}\end{figure}

In Fig.~\ref{menush03}(a) the experimental $\chi(k)k^2$ EXAFS for BaBiO$_3$
at 7\,K for the first Bi-O near-neighbour shell is shown. The pronounced
beating near 8\,\r{A}$^{-1}$ evidences for existence of at least two different
lengths of Bi-O bonds. In all previous EXAFS-researches
\cite{Balzarotti1,Boyce1,Heald1,Salem1,OurPaper4} the EXAFS-function was
quite successfully fitted in harmonic approximation as a sum of two
independent harmonic functions with different bond lengths and Debye-Waller
factors. However, temperature dependencies of Debye-Waller factors contradict
to the harmonic Einstein model \cite{OurPaper4}. This argues against the
independence of oxygen atom vibrations in two neighbouring octahedra.

In this work, we model the oscillatory Bi-O potential as follows. Inequivalence
of the two types of BiO$_6$ octahedra is due to presence or absence of a hole
pair in the hybridised molecular orbitals Bi$6s$O$2p_{\sigma^*}$ (see
Sec.~\ref{Relationship}). Suppose, the movement of oxygen atoms may transfer
a hole pair from one octahedron to another. Such a movement exchanges the roles
of two inequivalent octahedra and requires a double-well form of oscillatory
Bi-O potential. Here, we take a parabolic form of each well
$U_1=\kappa_1(r-r_1)^2/2$ and $U_2=\kappa_2(r-r_2)^2/2$ which are joined
continuously. Given the calculated $\chi(k)$, defined by Eqs.~(\ref{PRDF}) and
(\ref{gChi}), we perform a least-squares fit between the model and experimental
$\chi(k)$ over the range $k=$ 2--16\,\r{A}$^{-1}$ (see Fig.~\ref{menush03}(a)).
The six parameters determined by the fit were $E_{th}$, $r_1$, $r_2$,
$\kappa_1$, $\kappa_2$, and $N$. The number of parameters here is the same as
for fitting in the harmonic approximation.

The analysis of parameters of double-well potential allows one to draw the
following conclusions on the oscillations of oxygen atoms in BaBiO$_3$. (i)
At low temperatures there exist two explicit peaks on PRDF, that is why x-ray
and neutron diffraction detect the static distortions in BaBiO$_3$.
(ii) Even at low temperatures, the tunnelling probability between the two wells
is non-zero. As it will be discussed in Sec.~\ref{Relationship}, such a
tunnelling is equivalent to the dynamic exchange
Bi\underline{L}$^2$O$_6\leftrightarrow$ BiO$_6$ and explains observed
activation conductivity in BaBiO$_3$. (iii) The temperature rise leads to
gradual increase of probability of interwell tunnelling and to the structural
transition to cubic phase found at 750--800\,K \cite{Cox2}. (iv)
Non-equidistance of energy levels implies that oscillations of breathing- or
stretching types with several frequencies exist:  $\omega_0=E_1-E_0$,
$\omega_1=E_2-E_0$, $\omega_2=E_3-E_0$, etc. The latter are gradually
manifested at high temperatures. (v) The positions of minima of double-well
potential in our model are the {\it average} positions. They are spaced widely
at maximal octahedra tilting and spaced closely at minimal octahedra tilting.
In the latter case the probability of interwell tunnelling is maximal. In this
sense the tunnelling frequency $\omega_0$ is bounded by soft rotation mode
frequency.

\subsection{BaPbO$_3$}
\begin{figure}[!t]\begin{center}\includegraphics*[width=\hsize]{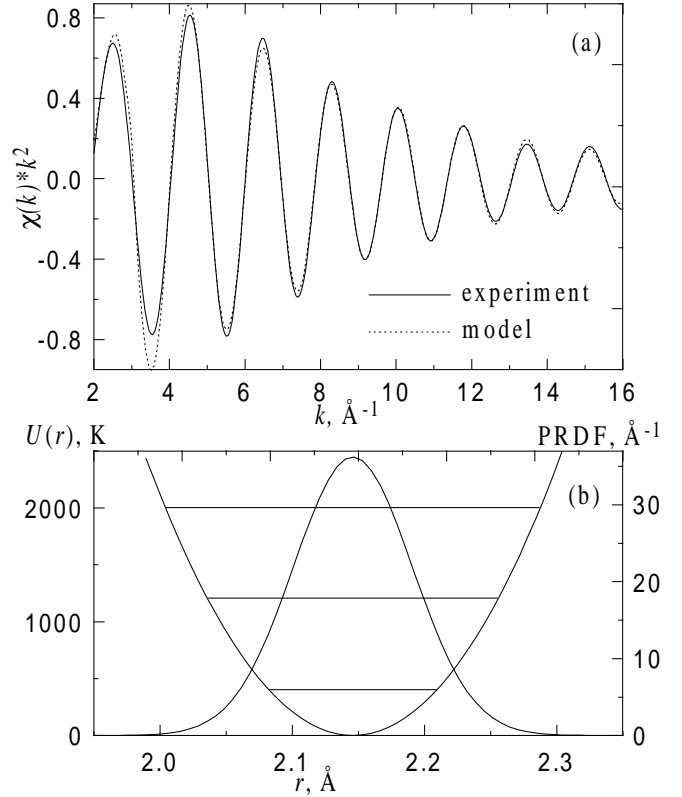}
\end{center}\caption{Experimental and model $\chi(k)k^2$ EXAFS for BaPbO$_3$
at 7\,K for the first Pb-O near-neighbour shell (a) and the model potential
with corresponding PRDF and energy levels (b).} \label{menush04}\end{figure}

The experimental $\chi(k)k^2$ EXAFS for BaPbO$_3$ for the first Pb-O
near-neighbour shell at all temperatures represents a sinusoid with no beatings
and phase breaks (Fig.~\ref{menush04}) and fitted well, using a single parabolic
potential. This means that all Pb-O bonds in BaPbO$_3$ are equivalent and the
breathing-type distortion is absent.

\subsection{Ba$_{1-x}$K$_x$BiO$_3$}

\begin{figure}[!t]\begin{center}\includegraphics*[width=\hsize]{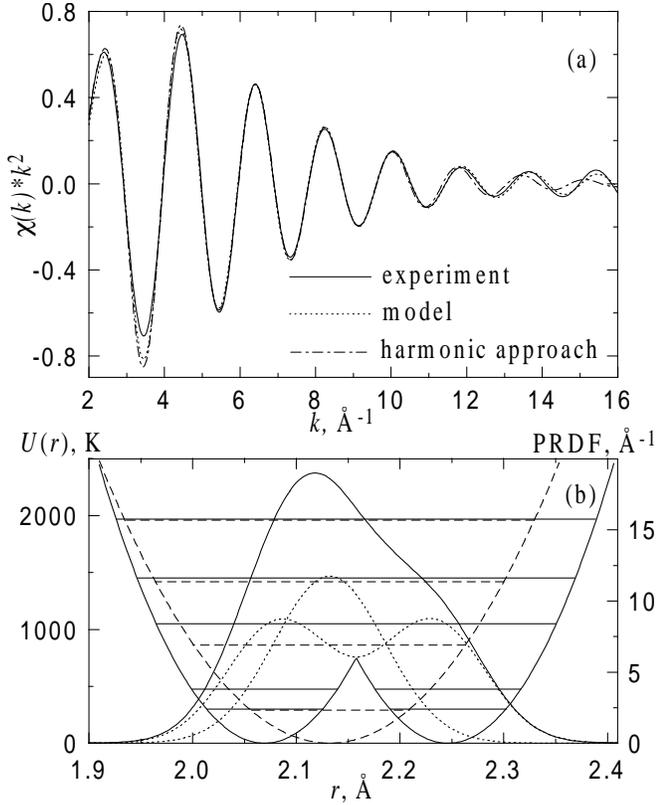}
\end{center}\caption{Experimental and model $\chi(k)k^2$ EXAFS for
Ba$_{0.6}$K$_{0.4}$BiO$_3$ at 7\,K for the first Bi-O near-neighbour shell (a)
and the model potentials (single-parabolic and double-parabolic) with
corresponding PRDF's and energy levels (b). Total PRDF is shown by solid line.}
\label{menush05}\end{figure}

The experimental $\chi(k)k^2$ EXAFS function for BKBO ($x=0.4$) at 7\,K for the
first Bi-O near-neighbour shell is shown in Fig.~\ref{menush05}(a), solid
curve.  Also, the model calculated in harmonic approach (as in
\cite{Heald1,Salem1}) is shown as a dash-dot curve. It is seen that
in the range $k\gtrsim12$\,\r{A}$^{-1}$ the harmonic approach fails. This argues
for the anharmonic Bi-O vibration behaviour similar to BaBiO$_3$ case also in
superconducting compositions BKBO. To a possible tendency in the cubic
superconducting materials towards the same type of distortions as in BaBiO$_3$
was pointed earlier \cite{Heald1,Boyce3}.

\begin{figure}[!t]\begin{center}\includegraphics*[width=\hsize]{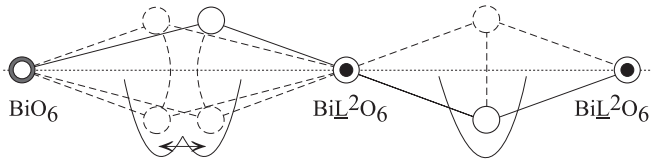}
\end{center}\caption{The sketch of oxygen vibrations in Ba$_{1-x}$K$_x$BiO$_3$
in the case of different (left) and equivalent (right) neighbouring octahedra
with corresponding radial oscillatory potentials.}\label{menush06}\end{figure}

The K doping of BaBiO$_3$ leads to partial replacement of the larger soft
octahedra BiO$_6$ by the smaller rigid octahedra Bi\underline{L}$^2$O$_6$ (see
details in Sec.~\ref{Relationship}). This causes a decrease and disappearance
of static breathing and tilting distortions, but keeps the different rigidities
of Bi-O bonds. Hence, the movement of an oxygen atom depends on that to which
neighbouring octahedra this atom belongs. If the neighbouring octahedra are
different, the oxygen atom oscillates in double-well potential as in
BaBiO$_3$. If the octahedra are equal, the oxygen atom oscillates in
single-parabolic potential as in BaPbO$_3$ (see Fig.~\ref{menush06}). The
statistical weights of these two cases depend on the potassium content $x$, and
are $(1-x)$ and $x$, correspondingly. Here, the force constants of the two
parabolas are assumed to be equal since their independent varying does not
improve the fits, and the number of fitting parameters equals 7. The resulting
model curve is presented in Fig.~\ref{menush05}(a), dotted. It is seen from
Fig.~\ref{menush05}(b) that the total PRDF is unsplit, so it is not surprising
that crystallographic measurements reveal the cubic structure \cite{Pei2}
(e.g. at $T=10$\,K $a=4.2742(1)$). Meanwhile, the first momentum of the
radial distribution function derived from our model is in a good agreement
with the diffraction data \cite{Pei2}, which validates our calculated amplitudes and
phases.

\begin{figure}[!t]\begin{center}\includegraphics*[width=\hsize]{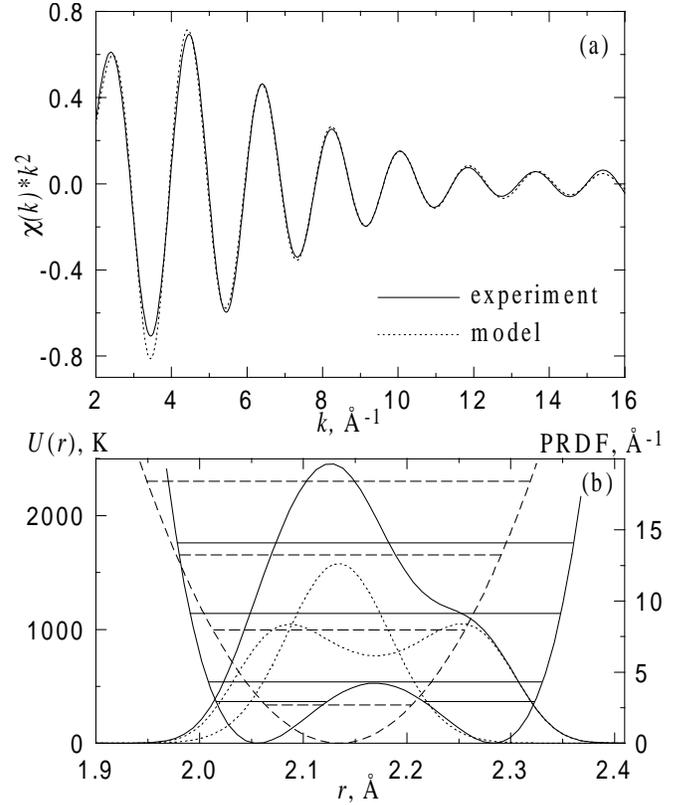}
\end{center}\caption{Experimental and model $\chi(k)k^2$ EXAFS for
Ba$_{0.6}$K$_{0.4}$BiO$_3$ at 7\,K for the first Bi-O near-neighbour shell (a)
and the model potentials (single-parabolic and polynomial of degree 4) with
corresponding PRDF's and energy levels (b). Total PRDF is shown by solid line.}
\label{menush07}\end{figure}

\begin{table*}[!t]
\center
\caption{Bi(Pb)-O parameters resulting from the fit to EXAFS data for
BaPbO$_3$ and Ba$_{1-x}$K$_x$BiO$_3$ with $x=$ 0, 0.4, 0.5. $r$ and $l$
are positions (\r{A}) of potential minima and corresponding PRDF's
first momenta, $\kappa$ is the force constant (eV/\r{A}$^2$). The index 0
corresponds to parabolic potentials, the indices 1 and 2 correspond to the two
parts of double-parabolic potentials.}
{\scriptsize
\begin{tabular}{l@{\ }l@{\ }r@{\quad}
*{4}{l@{\ }}*{2}{r@{\ }}@{\quad}
*{4}{l@{\ }}*{2}{r@{\ }}@{\quad}
*{4}{l@{\ }}*{2}{r@{\ }}}
&\multicolumn{2}{l}{BaPbO$_3$}&\multicolumn{6}{c}{$x=0$}&%
\multicolumn{6}{c}{$x=0.4$}&\multicolumn{6}{c}{$x=0.5$}\\
$T$,K&$r_0$=$l$&$\kappa_0$&
$r_1$&$r_2$&$l_1$&$l_2$&$\kappa_1$&$\kappa_2$&
$r_1$&$r_2$&$r_0$&$l$&$\kappa_{1,2}$&$\kappa_0$&
$r_1$&$r_2$&$r_0$&$l$&$\kappa_{1,2}$&$\kappa_0$\\
\hline
7  &2.15 &24&2.08 &2.30 &2.09 &2.285&11& 9&2.07 &2.245&2.13 &2.145&17& 9&2.065&2.20 &2.11 &2.12 &38&20\\
10 &2.15 &17&     &     &     &     &  &  &     &     &     &     &  &  &2.07 &2.20 &2.12 &2.125&27&11\\
20 &2.155&15&     &     &     &     &  &  &2.07 &2.235&2.13 &2.14 &28&18&2.05 &2.20 &2.12 &2.125&36&16\\
30 &     &  &2.09 &2.305&2.095&2.295&15&14&     &     &     &     &  &  &2.06 &2.195&2.12 &2.125&33&17\\
40 &2.145&11&     &     &     &     &  &  &2.07 &2.24 &2.12 &2.145&28&18&     &     &     &     &  &  \\
55 &     &  &2.09 &2.30 &2.10 &2.29 &16&14&2.075&2.25 &2.12 &2.15 &20&12&2.07 &2.20 &2.12 &2.125&41&22\\
65 &     &  &2.105&2.315&2.11 &2.305&16&13&2.08 &2.24 &2.13 &2.145&20& 9&2.06 &2.20 &2.12 &2.125&39&21\\
95 &2.14 &13&2.10 &2.31 &2.11 &2.30 &12& 7&     &     &     &     &  &  &2.06 &2.20 &2.125&2.13 &32&17\\
115&     &  &2.10 &2.31 &2.11 &2.30 &11& 6&2.08 &2.24 &2.12 &2.14 &18&10&     &     &     &     &  &  \\
135&     &  &2.105&2.31 &2.11 &2.30 &13& 5&     &     &     &     &  &  &2.07 &2.20 &2.12 &2.125&33&16\\
195&     &  &     &     &     &     &  &  &2.08 &2.24 &2.12 &2.145&20&11&     &     &     &     &  &  \\
300&2.14 & 9&2.10 &2.26 &2.11 &2.25 &11& 7&2.07 &2.23 &2.115&2.135&18& 9&2.07 &2.205&2.125&2.13 &24&11\\
\end{tabular}} \label{TabBiO}
\end{table*}

In this paragraph we address to the important question of statistical grounds
for the choice among several possible models. Consider, first, the statistical
chi-square function (labelled just like EXAFS-function, but this is the
different value)
\begin{equation}\label{chi2}
\chi^2=\frac{N_{exp}}{M}\sum_i^M
\left(\frac{\chi_{exp}(k_i)-\chi_{mod}(k_i)}{\varepsilon_i}\right)^2,
\end{equation}
where M is the number of data points in the fit, $N_{exp}$ was introduced
above, $\varepsilon_i$ are the individual errors in the experimental data
points. The latter were calculated as average of $\sqrt{1/I_0+1/I_t}$,
$I_0$ and $I_t$ being the intensities measured in the transmission EXAFS
experiment. The $\chi^2$ value must follow the $\chi^2$ distribution law with
degrees of freedom $\nu=N_{exp}-P$, where $P$ is the number of parameters varied
during the fit. That is $\chi^2$ must be less than the critical value
$X_{\nu}^c$ of the $\chi^2$ distribution with $\nu$ degrees of freedom and the
confidence level $c$. For our model in Fig.~\ref{menush05}(a) $\nu_2=11-7=4$ and
$\chi^2_2=5.3<X_4^{0.95}=9.5$, but for single-Gaussian model $\nu_1=11-4=7$ and
$\chi^2_1=16.8>X_7^{0.95}=14.1$. Thus, our model meets the $\chi^2$-test while
the simple harmonic model does not. It is quite natural that having increased
the number of parameters we got decreased $\chi^2$ value. But what the gain
should be? The comparison between the two models can be performed on the basis
of $F$-test. If the difference $\chi^2_1-\chi^2_2$ is physically meaningful,
not due to presence of the noise, that is if the simpler model cannot describe
some features in principle, this difference must {\it not} follow the $\chi^2$
distribution law with $\nu_1-\nu_2$ degrees of freedom. Provided that
$\chi^2_2$ follows the $\chi^2$ distribution with $\nu_2$ degrees of freedom,
the value $F(\nu_1-\nu_2,\nu_2)=(\chi^2_1/\chi^2_2-1)\nu_2/(\nu_1-\nu_2)$ must
not follow Fisher's $F$-distribution with $\nu_1-\nu_2$ and $\nu_2$ degrees of
freedom. That is $F$ must be greater than the critical value
$F_{\nu_1-\nu_2,\nu_2}^c$ of the $F$ distribution with $\nu_1-\nu_2$ and $\nu_2$
degrees of freedom and the confidence level $c$. Comparing the two models in
Fig.~\ref{menush05}(a), we find $F(3,4)=2.89$, which is equal to
$F_{3,\ 4}^{0.84}$. Therefore, we can claim with 84\% probability that we propose a
better model than the simplest single-Gaussian.

The complete results for Bi(Pb) octahedral ($N=6$) oxygen environment in
BaPbO$_3$ and Ba$_{1-x}$K$_x$BiO$_3$ with $x=$ 0, 0.4, 0.5 at various
temperatures are listed in Table~\ref{TabBiO}. The uncertainties in the EXAFS
distances and force constants are less than $\pm0.4\%$ and $\pm50\%$,
respectively. (These increments cause an increase of less than $10\%$ in the
value of misfit.) It is seen from Figs.~\ref{menush03} and \ref{menush05} that,
in general, the positions of potential minima are not equal to the positions of
PRDF's maxima. Because of that, among other parameters, we give the values of
PRDF's centers of gravity, i.e. values that can be defined
crystallographically. For BaBiO$_3$, the centers of gravity were calculated for
the two peaks of PRDF separately.

The parameters of potentials were obtained through the fit and, of course,
depend on the form of the model potentials. To elucidate this influence, we
constructed the model EXAFS function, using the polynomial of degree four:
$U=-\kappa(r-r_0)^2/2+\xi(r-r_0)^4+\kappa^2/(16\xi)$, where the last
term is introduced to zero the minimum values (Fig.~\ref{menush07}). The
number of fitting parameters here is the same as for double-parabolic
potential. It turned out, that for the both potential forms the mean
frequencies of Bi-O oscillations are practically identical for all
temperatures. Because of this, in the present work we use double-parabolic
potentials with parameters of clear meaning (potential minima positions and
force constants) rather than a polynomial potential with abstract coefficients.
Since the model EXAFS function weakly depends on the shape and value of
interwell energy barrier $U_0$, we can not determine the $U_0$ value from EXAFS
measurements exactly.

It worth to notice that in the case of BaBiO$_3$ the above model of oxygen
vibrations has some limitation due to the existence of the static rotation
(tilting) distortion $\sim11^\circ$. Thus, the distances observed from EXAFS
spectra treatment are measured along Bi-O-Bi zigzag curve, but not along the
Bi-Bi ([100]) direction as in BKBO. This may result in some difference between
$\kappa1$ and $\kappa2$ values (see Fig.~\ref{menush03}(b) and
Table~\ref{TabBiO}) and some deviation of calculated frequency from real ones in
BaBiO$_3$.

\section{Relationship between the local crystal and local electronic
structures}\label{Relationship}

\subsection{BaBiO$_3$}
The co-existence in BaBiO$_3$ of two different types of octahedra
with two different Bi-O bond lengths and strengths reflects the different
electronic structures of BiO$_6$ complexes.

\begin{figure}[!t]\begin{center}
\includegraphics*[width=\hsize]{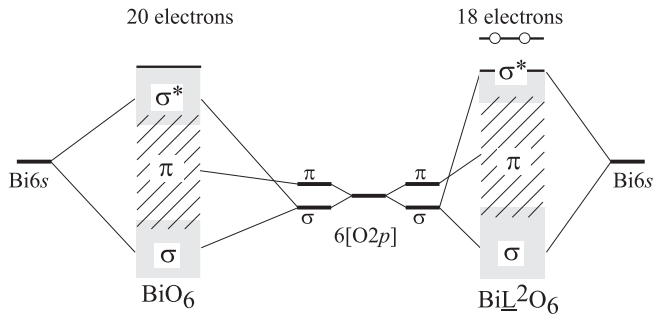}\end{center}
\caption{The scheme of electronic structure formation for two
different octahedral BiO$_6$ complexes.} \label{menush08}\end{figure}

Octahedral complexes represent the most tightly bound items of the
perovskite-like structures because of strong covalence of Bi$6s$-O$2p_\sigma$
bonds. The valence band structure of BaBiO$_3$ is determined by overlapping of
Bi$6s$ and O$2p$ orbitals \cite{Sleight1,Mattheiss1}, and, owing to strong
Bi$6s$-O$2p_\sigma$ hybridisation, the octahedra can be considered as
quasi-molecular complexes \cite{Sugai5}. Each complex has ten electron
levels consisting of a pair of bonding Bi$6s$-O$2p_\sigma$, six nonbonding
O$2p_\pi$, and a pair of antibonding Bi$6s$O$2p_{\sigma^*}$ orbitals. A unit
cell, which includes two octahedra, has 38 valence electrons (10 from two
bismuth ions, 4 from two barium ions, and 24 from six oxygen atoms). However,
the numbers of occupied states in the two octahedtral complexes are different:
octahedron Bi\underline{L}$^2$O$_6$ carries 18 electrons and has one free level
or a hole pair \underline{L}$^2$ in the upper antibonding
Bi$6s$O$2p_{\sigma^*}$ orbital, in octahedron BiO$_6$ with 20 electrons both
antibonding orbitals are filled (Fig.~\ref{menush08}). It is quite natural that
Bi\underline{L}$^2$O$_6$ octahedra have stiff (quasi-molecular) Bi-O bonds and
the smaller radius, and BiO$_6$ octahedra represent non-stable molecules with
filled antibonding orbitals and the larger radius. Because the sum of two
nearest octahedra radii overcomes the lattice parameter $a$, the octahedral
system must tilt around [110] axis, producing a monoclinic distortion in
BaBiO$_3$ (see Fig.~\ref{menush02}).

The assumption of equal electron filling for nearest octahedra
(Bi\underline{L}$^1$O$_6+$Bi\underline{L}$^1$O$_6$) contradicts to
experiments, since in this case equal Bi-O bond lengths and local magnetic
ordering should be observed.

Therefore our new scheme of bismuth disproportionation
2Bi\underline{L}$^1$O$_6\to$ Bi\underline{L}$^2$O$_6$+BiO$_6$ is in full
agreement with charge balance, presence of two types of octahedron complexes
and absence of any local magnetic moment \cite{Uchida2,Uemura1}.

Because Ba$^{2+}$ in perovskite-type lattice is bound by pure ionic bond, the
electron density is concentrated mainly in octahedra volume
$V_0$ \cite{Hamada1}. Hence, the energy of the highest occupied level $E_f$ is
related to the octahedron radius $R$ and the number of valence electrons $N$
in a unit cell as
\begin{equation}\label{E_f}
E_f\sim h^2N^{2/3}/m_eV_0^{2/3}\sim h^2N^{2/3}/m_eR^2,
\end{equation}
where $h$ is the Planck's constant, $m_e$ is the electron mass. This
qualitative relation leaves out of account the deviation of Fermi surface from
sphere \cite{Uchida1}. Nevertheless, the value of $E_f$, which transforms to
the Fermi level $E_F$ at spatial overlapping of equal octahedra in the crystal,
is in close connection with the octahedron radius and with the number of
valence electrons.

\begin{figure*}[!t]\begin{center}
\includegraphics*[width=\hsize]{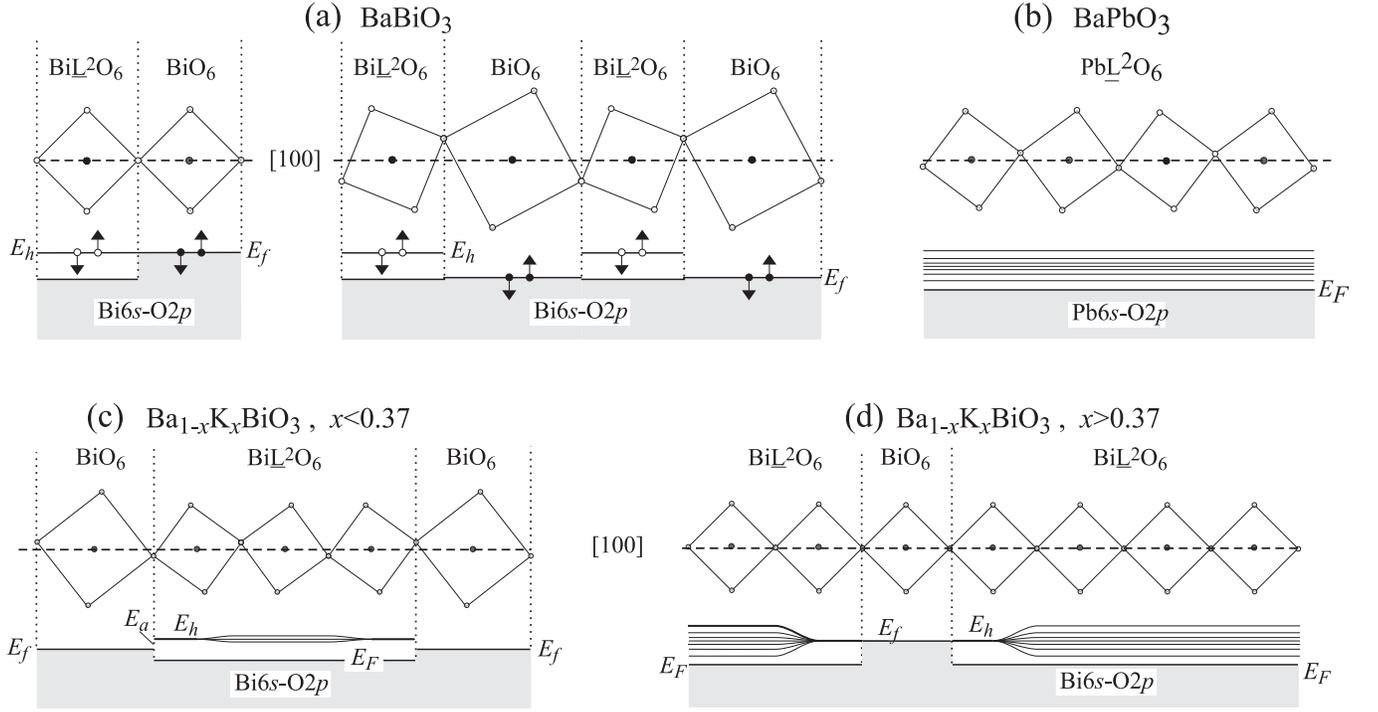}\end{center}
\caption{The scheme of relationship between the local crystal and local
electronic structures. (a) The ground state of BaBiO$_3$. Left --- in cubic
phase, right --- in monoclinic phase. $E_h$ is the energy of localised free
level, $E_f$ is the energy of the highest occupied Bi$6s$O$2p_{\sigma^*}$
orbital. The black and white circles with arrows denote electrons and holes,
correspondingly, with opposite spin orientations. The occupied states of
Bi$6s$O$2p$ valence band are marked by grey. At the top, the local structure of
octahedral complexes is shown. The [100] direction is equivalent to Bi-Bi
direction. (b) The ground state of BaPbO$_3$. The unoccupied band above the
Fermi level $E_F$ is shown. (c) The ground state of Ba$_{1-x}$K$_x$BiO$_3$,
$x<0.37$.  The splitting of free level $E_h$ at spatial overlapping of
Bi\underline{L}$^2$O$_6$ octahedra is sketched. $E_a$ is the activation energy
$E_a=E_h-E_f$. (d) The ground state of Ba$_{1-x}$K$_x$BiO$_3$, $x>0.37$. The
formation of unoccupied band above the Fermi level $E_F$ at the percolation
threshold reaching is sketched.} \label{menush09}\end{figure*}

According to expression (\ref{E_f}), the local electronic structure is
connected with the local crystal structure. In the left of
Fig.~\ref{menush09}(a), the case of hypothetical simple cubic structure of
BaBiO$_3$ is shown. Real monoclinic structure arises from combined breathing
(alternating octahedra with different radii $R$) and rotation distortions,
which, according to (\ref{E_f}), leads to the lowering of the energy $E_f$ of
the highest occupied Bi$6s$O$2p_{\sigma^*}$ orbital in BiO$_6$ octahedron in
comparison with energy $E_h$ of unoccupied Bi$6s$O$2p_{\sigma^*}$ orbital in
Bi\underline{L}$^2$O$_6$ octahedron. Hence, the repetitive sequence of empty
level $E_h$ on the background of fully filled valence band represents the
electronic structure of the ground state of BaBiO$_3$ (Fig.~\ref{menush09}(a),
right). In such a system there are no free carriers and conductivity occurs at
hopping of carrier pair from one octahedron to another with the activation
energy $E_a=E_h-E_f$. In this process the movement of hole pair in real space
leads to change of large octahedra to small ones and vice versa or to a dynamic
exchange Bi\underline{L}$^2$O$_6\leftrightarrow$ BiO$_6$, in full accordance
with vibrations in double-well potential showed in Fig.~\ref{menush03}, and
causes the hole-type conductivity. In this case, the activation energy $E_a$
means the pair localisation energy.

This picture agrees with results of investigation of temperature dependencies
of conductivity and Hall coefficient: $n(T)=n_0\exp(-E'_a/k_BT)$
\cite{Uchida1}, where the value of hole concentration
$n_0=1.1\times10^{22}$\,cm$^{-3}$ coincides with the concentration of unit
cells, and the activation energy $E'_a=0.24$\,eV, if one takes into account that
in the case of two-particle activation conductivity, the number of hole pairs
is equal to the number of Bi\underline{L}$^2$O$_6$ complexes (concentration
of which is half $n_0$), and the activation energy $E_a=2E'_a=0.48$\,eV. It
should be mentioned that the possibility of the two-particle (bipolaron)
conductivity as a most probable mechanism of conductivity in BaBiO$_3$ was
pointed to earlier \cite{Uchida1}.

Pair splitting and hopping of a single electron from one octahedron to another
cost energy and lead to electronic structure reconstruction of the both
octahedral complexes. Such a splitting can be achieved under optical
excitation, which was observed experimentally as a photoconductivity peak at
the photon energy $h\nu=1.9$\,eV \cite{Uchida1}. The optical gap $E_g$ is the
energy difference between the excited and ground states. In the excited state,
BaBiO$_3$ has the local lattice of equivalent octahedra
Bi\underline{L}$^1$O$_6$ and possesses the non-compensated spin. Admittedly, it
must have anti-ferromagnetic ordering, as the ground state in undoped cuprates
La$_2$CuO$_4$ and Nd$_2$CuO$_4$. Such optical excitation leads to the local
dynamical lattice deformation observed in Raman spectra as a breathing mode
$\sim$570\,cm$^{-1}$ of giant amplitude under resonant coinciding of photon
energy of Ar$^+$ laser with $E_g$ \cite{Sugai1,Tajima3,Sugai5}. Using
lasers with other quantum energies destroys this resonant effect and leads to
abrupt decrease of breathing mode amplitude \cite{Tajima3}.

Thus, the scheme proposed accounts for the nature of the two energy gaps in
BaBiO$_3$. The activation energy $E_a$ appears in transport measurements and
is connected with coherent delocalisation of hole pairs owing to dynamic
exchange Bi\underline{L}$^2$O$_6\leftrightarrow$ BiO$_6$. The optical gap is
radically differs from a traditional gap in semiconductors. It corresponds to
energy difference between the excited and ground states.

\subsection{BaPbO$_3$}
In BaPbO$_3$, each octahedron also has ten molecular orbitals, nine of which
are filled and one, antibonding Pb$6s$O$2p_{\sigma^*}$, is unoccupied as in
octahedron Bi\underline{L}$^2$O$_6$, since lead has one electron less than
bismuth. Thus, all the octahedral complexes represent the bound molecules
Pb\underline{L}$^2$O$_6$ with equal radii. At spatial overlapping of
Pb\underline{L}$^2$O$_6$ in crystal BaPbO$_3$, wave functions of
Pb$6s$O$2p_{\sigma^*}$ states become delocalised, unoccupied levels $E_h$ of
neighbouring octahedra split into a free carrier band, and, merging with the top
of valence band consisting of initially filled Pb$6s$O$2p_{\sigma^*}$ orbitals,
make the half filled conduction band (Fig.~\ref{menush09}(b)). It is important to
notice that unoccupied Bi(Pb)$6s$O$2p_{\sigma^*}$ orbitals in the case of
absence of spatial overlapping in BaBiO$_3$ behave as localised hole pairs, but
transform to the conduction band at the spatial overlapping in BaPbO$_3$.
Because of this, BaBiO$_3$ appears to be a semiconductor of $p$-type, and
BaPbO$_3$ --- a semimetal of $n$-type. Since the radius of
Pb\underline{L}$^2$O$_6$ complex is slightly greater than that of
Bi\underline{L}$^2$O$_6$ complex, the Fermi level $E_F$ in BaPbO$_3$,
according to expression (\ref{E_f}), lies lower in comparison with $E_f$ in
BaBiO$_3$, which agrees with results of photoelectron spectroscopy
\cite{Namatame2}.

Thus, the lead valence state Pb\underline{L}$^2$O$_6$ in BaPbO$_3$ is similar
to the bismuth state in Bi\underline{L}$^2$O$_6$ complexes in BaBiO$_3$.
According to Fig.~\ref{menush04}, oxygen atoms in BaPbO$_3$ oscillate in the
parabolic potential without any exchange between equal Pb\underline{L}$^2$O$_6$
octahedra. Therefore, there is no charge pair transfer in semimetallic
BaPbO$_3$ and the itinerant electrons behave as usual Fermi liquid.

\subsection{Ba$_{1-x}$K$_x$BiO$_3$}
The doping of BaBiO$_3$ by lead or potassium leads to decrease of integral
structural lattice distortions and causes essential changes in both local
crystal and electronic systems. Firstly, we discuss the changes in electronic
structure of BKBO as a simpler case, and then will extend the obtained
conclusions to the BPBO system.

Since K$^+$ ion has one valence electron instead of two of Ba$^{2+}$ ion,
the substitution of each two Ba$^{2+}$ for two K$^+$ ions produces an
additional unoccupied level or hole pair in a Bi$6s$O$2p_{\sigma^*}$ orbital
and modifies the BiO$_6$ complex to the Bi\underline{L}$^2$O$_6$. As a result,
the ratio of the numbers of Bi\underline{L}$^2$O$_6$ and BiO$_6$ complexes
changes from $1:1$ in BaBiO$_3$ to $(1+x):(1-x)$ and equals $7:3$ in
Ba$_{0.6}$K$_{0.4}$BiO$_3$ and $3:1$ in Ba$_{0.5}$K$_{0.5}$BiO$_3$. The spatial
overlapping of Bi\underline{L}$^2$O$_6$ complexes appear, which, taking into
account their small radii and rigid bonds, contracts the lattice. This leads
to decrease and disappearance (at $x>0.37$) of both static rotation- and
breathing-type distortions. The lattice is forced to contract, despite the
ionic radii of Ba$^{2+}$ and K$^+$ are practically equal.

Structural changes are accompanied by essential changes in physical properties
of BKBO: at $x\approx0.37$ the phase transition insulator-metal occurs and the
superconductivity arises that remains up to the dopant concentration $x=0.5$
corresponding to the solubility limit of potassium in BaBiO$_3$. The type
of the temperature dependence of conductivity changes from semiconducting to
metallic one, Hall coefficient changes its sign, and, in the normal state, BKBO
compound with $x>0.37$ behaves as a metal with $n$-type conductivity
\cite{Sato1}.

These changes are well described in the framework of the above scheme
(Fig.~\ref{menush09} for BKBO). At the low dopant amount ($x<0.37$), the
contraction of the larger (soft) and the stretching of the smaller (rigid)
octahedra bring, according to expression (\ref{E_f}), $E_h$ and $E_f$
energies close together and the activation energy $E_a=E_h-E_f$ decreases. As
dopant concentration rises, the number of Bi\underline{L}$^2$O$_6$ octahedra
increases, and at $x>0.37$, in the lattice arise the three-dimensional chains
of spatially overlapped Bi\underline{L}$^2$O$_6$ octahedral complexes
\cite{percol}, their unoccupied levels split and form the conductivity band,
which is equivalent to the percolation threshold reaching that determines the
insulator-metal phase transition. Here, the itinerant electrons show Fermi
liquid behaviour with Fermi level $E_F$ as in BaPbO$_3$. However, in contrast to
the picture in BaPbO$_3$, in the BKBO structure there exist complexes BiO$_6$
through which itinerant electrons can not move because all the levels in these
complexes are occupied. The movement of electrons occupying the upper
Bi$6s$O$2p_{\sigma^*}$ orbital is possible only in pairing state at the dynamic
exchange Bi\underline{L}$^2$O$_6\leftrightarrow$ BiO$_6$, since the unpairing
costs energy consumption. But, in contrast to the case of BaBiO$_3$, the static
rotation- and breathing-type distortions of the octahedra disappear, their mean
radii become approximately equal (and equal to half the lattice parameter), and
the pair localisation energy approaches zero: $E_a=E_h-E_f\to0$
(Fig.~\ref{menush09}(d)). In this case, the delocalised carrier pairs can freely
move through the octahedral system that explains the appearance of
superconductivity in BKBO ($x>0.37$). This scheme of the local electronic
structure is in full agreement with the picture of local oxygen vibration in
double-well potential (see Figs.\ref{menush05},\ref{menush06},\ref{menush07}).

The unpairing of electrons, as in BaBiO$_3$, is possible under optical
excitation and is manifested by a pseudo-gap observed in reflectivity spectra
even in the metallic phase of BPBO-BKBO systems. For instance, in BKBO with
$x=0.4$ the pseudo-gap is about 0.5\,eV \cite{Blanton1} and about 0.6\,eV
\cite{Uchida1} in BPBO with $x=0.25$.

Our model suggests that in the metallic phase two carrier types are
present: itinerant electrons and pairs of initially (in BaBiO$_3$) localised
carriers. The co-existence of the two carrier types was confirmed by
experiments on investigation of conductivity, Hall effect, and thermoelectric
power \cite{Uchida1}, as well as by photoemission spectra
\cite{Nagoshi2,Namatame2} and Raman spectra investigations
\cite{Sugai1,Sugai2}. These results are in a good agreement with observed
zero-bias conductance \cite{Hellman3}. The change in carriers behaviour from
localised to itinerant at doping of BaBiO$_3$ by potassium was pointed to in
x-ray absorption spectra analysis at the O $K$ edge \cite{Salem1}.
The analysis of EPR spectra \cite{Misra1,Yakubovskii1} showed the presence of
the localised carrier pairs, which also was confirmed by the observation of
two-particle tunnelling in the normal-state of BKBO \cite{Hellman3}.

\subsection{BaPb$_{1-x}$Bi$_x$O$_3$}

Practically the same changes in electronic structure arise at the doping of
BaBiO$_3$ by lead. The electronic structure of octahedral
Pb\underline{L}$^2$O$_6$ complex is entirely equivalent to that of
Bi\underline{L}$^2$O$_6$ complex. Unoccupied levels of overlapping octahedra in
BaPbO$_3$ form the conduction band, as at overlapping of
Bi\underline{L}$^2$O$_6$ in BKBO. At the doping of BaPbO$_3$ by bismuth
($0<x<0.37$), the case is similar to BKBO ($x>0.37$). However, in BPBO the
following combinations of neighbouring octahedra are possible:
Pb\underline{L}$^2$O$_6$-Pb\underline{L}$^2$O$_6$,
Pb\underline{L}$^2$O$_6$-Bi\underline{L}$^2$O$_6$,
Pb\underline{L}$^2$O$_6$-BiO$_6$, Bi\underline{L}$^2$O$_6$-BiO$_6$. This leads
to different local shifts of $E_h$ and $E_f$ energies, depending on different
octahedra neighbouring pairs. At the further increase of bismuth content, the
number of filled BiO$_6$ octahedra rises, the tilting and breathing distortions
enlarge, the pair localisation energy raises, the spatial overlapping of
unoccupied levels in Pb\underline{L}$^2$O$_6$ octahedra disappears, which
destroys the metallic type of conductivity and the system becomes a
semiconductor with $p$-type conductivity.

Thus, superconductive properties of BPBO in our model are connected with
dynamic exchange Pb\underline{L}$^2$O$_6\leftrightarrow$ BiO$_6$. Unfortunately,
it is practically impossible to observe the double-well-potential oxygen
vibrations for superconducting BPBO compositions ($0<x<0.35$) by EXAFS study
because of overlapping of Pb $L_3$ and Bi $L_3$ edges. Though an analysis of
EXAFS spectra for these compounds is possible under elaborate treatment
procedure \cite{Boyce2}, the precise enough values can be obtained only for
interatomic distances, but not for amplitude factors. For this reason, we do
not present the EXAFS data of BPBO ($0<x<1$) in this paper.

\subsection{Photoemission spectra anomalies}

\begin{figure}[!t]\begin{center}\includegraphics*[width=\hsize]{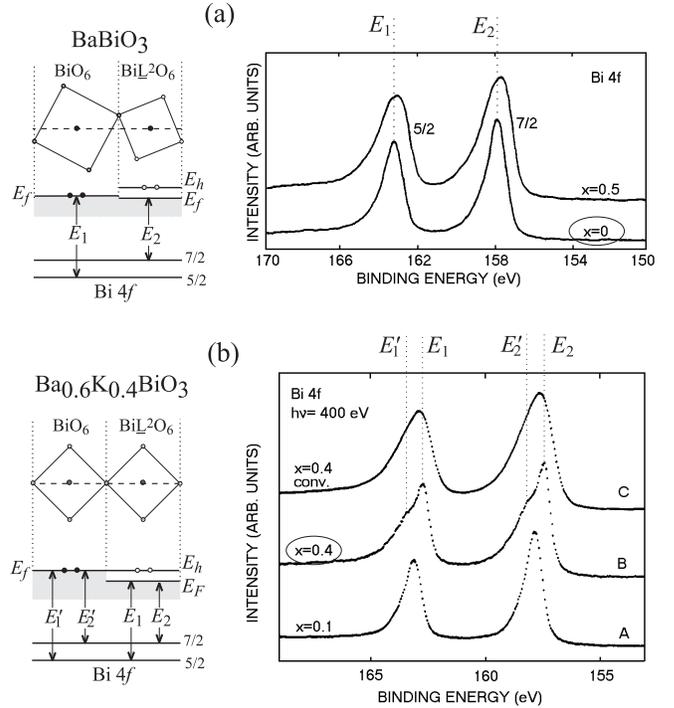}
\end{center}\caption{Binding energies of Bi$4f$(5/2, 7/2) core levels
in Ba$_{1-x}$K$_x$BiO$_3$: (a) for $x=0$, right --- experimental
photoemission spectra from \protect\cite{Nagoshi2} with energy
resolution $\Delta E=0.45$\,eV; (b) for $x=0.4$, right ---
experimental photoemission spectra from \protect\cite{Qvarford1}
with energy resolution $\Delta E=0.25$\,eV. Curve C is the result of
convolution of the curve B with Gaussian function with 0.75\,eV width.}
\label{menush10}\end{figure}

The analysis of photoemission spectra has revealed a serious contradiction
concerning the experimental observation of the splitting of the Bi$4f$(5/2, 7/2)
spectral lines in superconducting compounds BKBO \cite{Qvarford1} at the absence
of any peculiarities of those lines in parent BaBiO$_3$ \cite{Nagoshi2}. The
contradiction consists in that that though the analysis of x-ray diffraction
and EXAFS data points to existence of two different Bi-O bonds in BaBiO$_3$,
no splitting of Bi$4f$ doublet lines was observed, which indicates the
absence of considerable difference in bismuth valence states. The doping of
BaBiO$_3$ by potassium relieves the monoclinic distortion and equalises the
Bi-O bond lengths up to $a/2$. However, the Bi$4f$ spectral lines become
broaden \cite{Nagoshi2} and, at measurements on a high resolution spectrometer,
even split \cite{Qvarford1}, which, in contrast with simple cubic lattice in
BKBO ($x>0.37$), points to existence of two different bismuth valence states.

The above scheme of local electronic structure completely resolves this issue.
Indeed, in BaBiO$_3$ the binding energies of the Bi$4f$ core levels in
different Bi\underline{L}$^2$O$_6$ and BiO$_6$ octahedra are almost equal and
hence, the Bi$4f$ doublet lines in photoemission spectra are unsplit
(Fig.~\ref{menush10}(a)). Contrary to that, in Ba$_{0.6}$K$_{0.4}$BiO$_3$, the
binding energies of the Bi$4f$ core levels in the different octahedra $E'_1$,
$E'_2$ and $E_1$, $E_2$ differ on $\Delta E=E_h-E_F\approx E_a$, which causes
the broadening (see curve $x=0.5$ in Fig.~\ref{menush10}(a) from
\cite{Nagoshi2} and splitting (see curve $x=0.4$ in Fig.~\ref{menush10}(b) from
\cite{Qvarford1} of the Bi$4f$ doublet lines. The value of the splitting
coincides well with our estimation of $E_a=0.48$\,eV, and the ratio of
intensities $I$ of the core-level peaks corresponding to the binding energies
in different octahedral complexes BiO$_6$ and Bi\underline{L}$^2$O$_6$ is in
accordance with the relative content of these complexes in
Ba$_{0.6}$K$_{0.4}$BiO$_3$:  $I(E'_1)/I(E_1)\simeq I(E'_2)/I(E_2)\simeq3/7$.

\section{On possible nature of superconducting state}
\label{supercon}

Essentially anharmonic character of oxygen ion oscillations in soft rotation
mode is the base of anharmonic model of the high temperature superconductivity
\cite{Plakida1}. It was shown that anharmonic coupling constant $\lambda_s$
exceeds the constant in harmonic approach $\lambda_{ph}$:
\begin{equation}\label{lambdaPl}
\lambda_s/\lambda_{ph}\simeq J_s^2d^2\bar\omega/J_{ph}^2\langle
u^2\rangle\omega_s\gg1,
\end{equation}
which explains high critical temperatures in cuprate HTSC compounds as well as
in BPBO-BKBO systems. Here, $d$ and $\omega_s$ are the amplitude and frequency
of oscillation in a double-well potential, $\langle u^2\rangle$ is the
mean-square displacement of the ions with the mean frequency $\bar\omega$ in
harmonic approach, $J_{ph}^2$ and $J_s^2$ are, respectively, the deformation
potentials of harmonic and anharmonic oscillations averaged on the Fermi
surface.

In that model, the oscillatory movement of oxygen ions in rotation mode with a
large amplitude in the direction perpendicular to the Bi(Pb)-O bonds was
considered. A considerable gain in value of the coupling constant in expression
(\ref{lambdaPl}) is due to low frequency of soft rotation mode
($\omega_s<\bar\omega$) and the excess of its amplitude over amplitude of
harmonic oscillations ($d^2\gg\langle u^2\rangle$).

Meanwhile, previous \cite{Mattheiss1,Hamada1,Shirai1} as well as recent
(\cite{Meregalli1} and references therein) electronic structure
calculations showed that such a movement hardly affects the Bi-O bond lengths,
keeping $sp(\sigma)$ nearest-neighbour interaction nearly constant. Thus,
despite the existence of double-well potential in rotation (tilting) mode,
Meregalli and Savrasov \cite{Meregalli1} obtained very small anharmonic
contribution to $\lambda$. Also, they assumed the breathing-type oxygen
vibrations to be harmonic with small amplitude and high frequency and found too
small $\lambda\approx0.3$ to explain high $T_c$ in BKBO.

Our investigation firstly showed that in BKBO the double-well vibrations with
strong deformation potential also exist along the Bi-O-Bi direction. Such
vibrations have the breathing-like character and low frequency bounded by soft
rotation mode. Thus, our results resolve the above issue and explain the reason
for strong electron-phonon coupling in Bi-based oxides. Also, to these
vibrations we can assign the low-frequency part, below 40\,meV, of phonon
density of states observed in neutron scattering for superconducting BKBO
\cite{Loong2} and absent in calculations \cite{Meregalli1}. Besides, the
anomalous phonon softening along [100] direction \cite{Braden2} becomes clear
due to our observation of coherent breathing-like vibrations along [100] axis.

Hardy and Flocken \cite{Hardy1} evaluated values of $\lambda$, using abstract
model double-well potentials:
\begin{equation}\label{lambdaH}
\lambda(T)=N(0)\sum_{kk'}^{\rm FS}\sum_{n'>n}\frac%
{|\langle n|M_{kk'}|n'\rangle|^2}{E_{n'}-E_n}(f_n-f_{n'}),
\end{equation}
where $N(0)$ is the density of electron states at the Fermi level, $M_{kk'}$
is the {\it e}-ph matrix element between electronic states $|k\rangle$ and
$|k'\rangle$ on the Fermi surface, $|n\rangle$ and $|n'\rangle$ are the
oscillatory states with energies $E_n$ and $E_{n'}$, $f_n$ and $f_{n'}$ are the
thermal weighting factors; $f_n=\exp(-E_n/kT)/\sum_{n'}\exp(-E_{n'}/kT)$.

\begin{figure}[!t]\begin{center}\includegraphics*[width=\hsize]{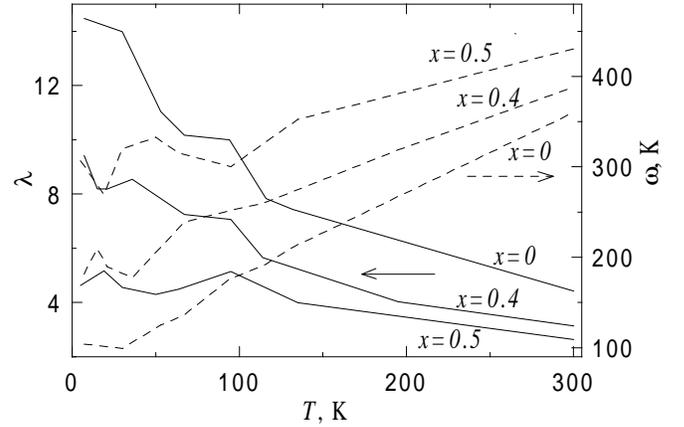}
\end{center}\caption{$\lambda$'s in terms of arbitrary multiplicative constant
(solid lines) and mean oscillatory frequencies (dashed lines) for
Ba$_{1-x}$K$_x$BiO$_3$ with $x=$ 0, 0.4, 0.5 for double-well potentials
obtained from EXAFS spectra treatment.} \label{menush11}\end{figure}

Given the oscillatory energies from our EXAFS-experiment, we have calculated
$\lambda$ in terms of arbitrary multiplicative constant which is proportional
to $N(0)$ and $M_{kk'}$ (Fig.~\ref{menush11}). Although ``phonon'' part of
$\lambda$ is the strongest for BaBiO$_3$, this composition is not
superconducting. On the one hand, because of pair localisation energy $E_a$. On
the other, the rotation oscillations of rigid Bi\underline{L}$^2$O$_6$
octahedra and ordinary phonons (of stretching and bending types) differ in
collective character of motion in BiO$_2$ planes. In BaBiO$_3$, rigid
Bi\underline{L}$^2$O$_6$ octahedra are separated by soft BiO$_6$ octahedra and
are not spatially overlapped. The partial replacement of the larger soft
octahedra by the smaller rigid ones at the potassium doping leads to decrease
and disappearance of localisation energy $E_a$ (Fig.~\ref{menush09}(c,d)) and to
spatial coherence of rotation oscillations with length of several lattice
parameters, depending on the doping level $x$. For superconducting
compositions, $T_c$ means the temperature up to which $\lambda$ is considerably
decreased (Fig.~\ref{menush11}) and/or the coherence in oscillations of
neighbouring octahedra is thermally destroyed by ordinary phonons.

The ground state of BaBiO$_3$ can be considered as a bipolaronic
state as well, and the movement of carrier pairs correlated with oxygen atoms
oscillations at the dynamic exchange
Bi\underline{L}$^2$O$_6\leftrightarrow$ BiO$_6$ evidences for possible
application of bipolaron theory \cite{Alexandrov}, which is supported by the
small size of the pair (octahedron size) and by the existence of the pair
state above $T_c$.

\section{Conclusion}\label{Conclusion}

From the EXAFS investigation of BPBO-BKBO systems, the local crystal structure
peculiarities connected with non-equivalence of
Bi\underline{L}$^2$O$_6$-BiO$_6$ octahedra were observed. It was found that
oxygen vibrations are well described using double-well potentials, which leads
to the strong electron-phonon coupling due to coherent modulation of Bi-O
bond lengths with low frequency and causes correlated carrier pair movement
with oxygen ions oscillations. The underlying relationship between the local
crystal and local electronic structures was established that explains the full
list of unusual properties of BPBO-BKBO systems: the existence of two energy
gaps (transport $E_a$ and optical $E_g$) in BaBiO$_3$; the mechanism of
two-particle conductivity in BaBiO$_3$; the co-existence of two different
carrier types; pseudo-gap observation in metallic compositions of BKBO and
BPBO; the observation of localised pairs from EPR spectra; the observation of
low-frequency ($<40$\,meV) phonons; the nature of concentration and temperature
phase transitions; the contradictions between local (Raman, and EXAFS) and
integral structural methods for metallic BKBO; the anomalies of XPS spectra in
BaBiO$_3$ and BKBO. We observed the correlated movement of carrier pairs and
low-frequency breathing-like oxygen octahedra vibrations, which corresponds to
superconducting state. The model proposed combines some principal features of
the real-space pairing \cite{Rice1}, anharmonic models \cite{Plakida1,Hardy1}
and bipolaron theory \cite{Alexandrov} of high-$T_c$ superconductivity.

The likeness of rotation mode peculiarities of BiO$_6$ octahedra and CuO$_n$
complexes and the anomalies in temperature dependencies of
Debye-Waller factors in cuprates \cite{XAFS_b} allow one to hope that
similar model approach can be applied to the Cu-based superconductors.

\acknowledgments
We acknowledge LURE Program Committee for beamtime providing, Professors
S.~Benazeth and J.~Purans for help during x-ray absorption
measurements. We are also grateful to Professor A.~P. Rusakov for high-quality
BKBO samples and to Dr.  A.~V. Kuznetsov and Dr. A.~A. Ivanov for helpful
discussions. The work is supported by RFBR (Grant No.  99-02-17343) and Program
``Superconductivity'' (Grant No. 99010).


\begin{thebibliography}{99}
\bibitem{percol}
For the simple 3D case (levels 0 and 1 allowable, percolation through the cubic
cell side) the percolation threshold is well known $p_c\approx0.31$. If the
ground state represents the alternating cells of 0 and 1 states, our
Monte Carlo calculations yield $p_c\approx0.15$. If we additionally require
that the percolating cluster must have at least one straight continuous chain
in chosen direction, the percolation threshold $p_c\approx0.37$.
\end{thebibliography}
\end{document}